\documentclass[prd,onecolumn,showpacs,preprintnumbers,amsmath,amssymb,floatfix,nofootinbib]{revtex4}
\usepackage{epsfig,epsf}
\usepackage{amsmath}
\usepackage{amsthm}
\usepackage{amsfonts}
\usepackage{amssymb}
\usepackage{dsfont}
\usepackage{multirow}
\usepackage{appendix}
\usepackage{slashed}
\usepackage[active]{srcltx}
\usepackage{psfrag}
\usepackage{multirow}
\def\beq{\begin{equation}}
\def\eeq{\end{equation}}
\def\bea{\begin{eqnarray}}
\def\eea{\end{eqnarray}}
\def\beeq{\begin{eqnarray}}
\def\eeeq{\end{eqnarray}}

\def\ba{\begin{array}}
\def\ea{\end{array}}

\def\xis0{{\Xi^{*0}}}

\def\g5{\gamma_5}

\def\baeq{\begin{appeq}}     \def\eaeq{\end{appeq}}  
\def\baeeq{\begin{appeeq}}   \def\eaeeq{\end{appeeq}}
\newenvironment{appeq}{\beq}{\eeq}   
\newenvironment{appeeq}{\beeq}{\eeeq}

\renewcommand{\theequation}{\arabic{equation}}
\newcounter{hran}
\renewcommand{\thehran}{\thesection.\arabic{hran}}

\def\bmini{\setcounter{hran}{\value{equation}}
\refstepcounter{hran}\setcounter{equation}{0}
\renewcommand{\theequation}{\thehran\alph{equation}}\begin{eqnarray}}
\def\bminiG#1{\setcounter{hran}{\value{equation}}
\refstepcounter{hran}\setcounter{equation}{-1}
\renewcommand{\theequation}{\thehran\alph{equation}}
\refstepcounter{equation}\label{#1}\begin{eqnarray}}

\newskip\humongous \humongous=0pt plus 1000pt minus 1000pt

\begin{document}

\preprint{}
\preprint{}
\title{ $\Lambda_b(6146)^0$ state newly observed by LHCb}

\author{K.~Azizi}
\affiliation{Department of Physics, University of Tehran, North Karegar Avenue, Tehran
14395-547, Iran}
\affiliation{Department of Physics, Do\u gu\c s University,
Ac{\i}badem-Kad{\i}k\"oy, 34722 Istanbul, Turkey}
\author{Y.~Sarac}
\affiliation{Electrical and Electronics Engineering Department,
Atilim University, 06836 Ankara, Turkey}
\author{H.~Sundu}
\affiliation{Department of Physics, Kocaeli University, 41380 Izmit, Turkey}

\date{\today}

\begin{abstract}
We study the bottom $\Lambda_b(6146)^0$ baryon, newly discovered by the LHCb Collaboration. By adopting an interpolating current of $(L_{\rho}, L_{\lambda})=(0,2)$ type and $D$-wave nature with spin-parity quantum numbers $J^P=\frac{3}{2}^+$ for this  heavy bottom baryon, we calculate its mass and residue. Using these spectroscopic parameters, we also investigate its dominant decays $\Lambda_b(6146)^0\rightarrow\Sigma_b\pi$ and  $\Lambda_b(6146)^0\rightarrow\Sigma^*_b\pi$ and estimate the width of $\Lambda_b(6146)^0$ obtained via these channels. The obtained mass, $m_{\Lambda_b}=(6144\pm 68)$~MeV is in accord nicely with the experimental data. The width obtained via the dominant channels is also consistent with the experimental data of LHCb collaboration.  We calculate the spectroscopic parameters and the same decay channels for the $c$-partner of $\Lambda_b(6146)^0$ state, namely  $\Lambda_c(2860)^+$, as well.  We compare the obtained results with the existing theoretical predictions as well as experimental data. The results indicate that the state $\Lambda_b(6146)^0$ and its charmed-partner $\Lambda_c(2860)^+$ can be considered as $1D$-wave baryons with $J^P=\frac{3}{2}^+$.
\end{abstract}

\maketitle
\section{Introduction}

The heavy baryons containing a heavy quark play an important role in our understanding of the strong interaction. Their quark content makes them more attractive in point of studying the dynamics of light quarks when a heavy one is present. The studies on the heavy baryons with one heavy quark could improve our understanding of the confinement mechanism and provide us with test of the quark model and heavy quark symmetry. And also, the investigations on their different properties could help  us test the predictions obtained by different theoretical assumptions on their internal organizations. Therefore, understanding the nature and  properties of these baryons and their quantum numbers by means of theoretical and experimental studies are of great importance.

In the last decades, the advances in experimental facilities and techniques led to the observations of many new states. The new observations include the conventional hadrons and the exotic states. Some of the baryons with single heavy quark content are among these states. In the Particle Data Group (PDG) listing~\cite{Tanabashi18} there exist seven $\Lambda_c$ states, which are $\Lambda_c^+$, $\Lambda_c(2595)^+$, $\Lambda_c(2625)^+$, $\Lambda_c(2765)^+$ (or $\Sigma_c(2765)$), $\Lambda_c(2860)^+$, $\Lambda_c(2880)^+$ and $\Lambda_c(2940)^+$. On the other hand, there are a smaller number of listed $\Lambda_b$ states, which are $\Lambda_b^0$,  $\Lambda_b(5912)^0$ and $\Lambda_b(5920)^0$. Among these states, the $\Lambda_{c}(2860)^+$ was discovered in 2017 by the LHCb Collaboration~\cite{Aaij:2017vbw}. Besides the first observation of this resonance by means of an amplitude analysis of $\Lambda_b\rightarrow D^0p\pi^-$ decay, the spin of $\Lambda_{c}(2880)^+$, which was firstly reported by the CLEO Collaboration~\cite{Artuso:2000xy}, was also confirmed in this work. The quantum numbers of the $\Lambda_{c}(2860)^+$ state were reported as $J^P=3/2^+$ and its measured mass and decay widths were presented as $m_{\Lambda_{c}(2860)^+}=2856.1^{+2.0}_{-1.7}\mbox{(stat)}\pm0.5\mbox{(syst)}^{+1.1}_{-5.6}\mbox{(model)}$~MeV and $\Gamma_{\Lambda_{c}(2860)^+}=67.6^{+10.1}_{-8.1}\mbox{(stat)}\pm1.4\mbox{(syst)}^{+5.9}_{-20.0}\mbox{(model)}$~MeV~\cite{Aaij:2017vbw}, respectively. Recently, the LHCb collaboration announced the observation of two bottom baryons with very close masses, which were reported as $m_{\Lambda_{b}(6146)^0}=6146.17\pm0.33\pm0.22\pm 0.16$~MeV and $m_{\Lambda_{b}(6152)^0}=6152.51\pm0.26\pm0.22\pm 0.16$~MeV. Their respective widths are $\Gamma_{\Lambda_{b}(6146)^0}=2.9\pm1.3\pm0.3$~MeV and $\Gamma_{\Lambda_{b}(6152)^0}=2.1\pm0.8\pm0.3$~MeV. According to their masses and widths, they were interpreted as a $\Lambda_b(1D)^0$ doublet~\cite{Aaij:2019amv}.

The properties of the heavy baryons with single heavy quark were studied by different approaches in the literature. Among some of these studies, including analyses on their mass spectrum or decay mechanisms, are the various quark models~\cite{Copley:1979wj,Maltman:1980er,Capstick:1986bm,Ivanov:1998wj,Ivanov:1999bk,Hussain:1999sp,Ebert:2005xj,Albertus:2005zy,Garcilazo:2007eh,Zhong:2007gp,Ebert:2007nw,Valcarce:2008dr,Roberts:2007ni,Ebert:2011kk,Hernandez:2011tx,Liu:2012sj,Karliner:2015ema,Yoshida:2015tia,Shah:2016mig,Shah:2016nxi,Chen:2016iyi,Thakkar:2016dna,Wang:2017kfr,Chen:2018vuc,Wang:2018fjm,Nagahiro:2016nsx,Yao:2018jmc,Wang:2019uaj}, relativistic flux tube model~\cite{Chen:2014nyo}, heavy hadron chiral perturbation theory~\cite{Huang:1995ke,Banuls:1999br,Cheng:2006dk,Cheng:2015naa,Jiang:2015xqa}, QCD sum rule method~\cite{Zhu:2000py,Wang:2010it,Mao:2015gya,Chen:2016phw,Mao:2017wbz,Wang:2017vtv,Aliev:2018vye,Aliev:2018lcs,Cui:2019dzj}, light cone QCD sum rules~\cite{Zhu:1998ih,Wang:2009ic,Wang:2009cd,Aliev:2009jt,Aliev:2010yx,Aliev:2014bma,Aliev:2016xvq,Chen:2017sci,Agaev:2017nn}, $^3P_0$ model~\cite{Chen:2007xf,Ye:2017dra,Ye:2017yvl,Yang:2018lzg,Chen:2017aqm,Guo:2019ytq,Lu:2019rtg,Liang:2019aag}, Bethe-Salpeter formalism~\cite{Guo:2007qu}, lattice QCD~\cite{Padmanath:2013bla,Bali:2015lka,Bahtiyar:2015sga,Bahtiyar:2016dom} and the bound state picture~\cite{Chow:1995nw}, etc. One may find more discussions about the related studies on the singly heavy baryons in the Refs.~\cite{Richard:1992uk,Korner:1994nh,Klempt:2009pi,Crede:2013sze,Cheng:2015iom,Chen:2016spr} and the references therein. 

In this work, we direct our attention to $1D$-wave charmed and bottom baryons with spin-$\frac{3}{2}$. Although our main focus is the bottom baryon $\Lambda_{b}(6146)^0$ that was recently observed by the LHCb Collaboration~\cite{Aaij:2019amv}, we also consider its charmed counterpart, $\Lambda_{c}(2860)^+$. We represent these two states as $\Lambda_Q$ where $Q$ is used to represent either $b$ or $c$ quark. Considering the proper interpolating currents for the considered states with quantum numbers $(L_{\rho}, L_{\lambda})=(0,2)$, we calculate the masses and the current coupling constants for these states using QCD sum rule approach~\cite{Shifman:1978bx,Shifman:1978by,Ioffe81}. The QCD sum rule method is a powerful nonperturbative method, which has provided  successful predictions for spectroscopic and decay properties of the hadrons, so far.  The $D$-wave charmed baryons were analyzed via the QCD sum rules in Refs.~\cite{Chen:2016phw,Wang:2017vtv}. In Ref.~\cite{Chen:2016phw}, both the charmed baryons and the bottom ones were considered in the framework of heavy quark effective theory. Ref.~\cite{Wang:2017vtv} presented the mass results  only for the charmed ones obtained in full QCD. In our case, we shall consider both the bottom and charmed baryons with light $u$ and $d$ quark content in full QCD. In the calculations, we adopt an interpolating current for the $\Lambda_{b}$ state considering the suggestion of the LHCb Collaboration as its possibly being one of $1D$ doublet of $\Lambda_{b}$ states. This suggestion was made considering the consistency of the mass of the observed $\Lambda_{b}$ states with the predictions presented by the constituent quark model~\cite{Chen:2014nyo,Capstick:1986bm}. Such spectroscopic analyses improve our understanding of the nature and structure of this baryons and contribute to our understanding of the nonperturbative natures of the strong interaction. From the analyses, we may deduce information about the quantum numbers of these states, as well. Beside these, another issue in baryon physics is the so-called missing resonances problem. According to the quark model, three constituent quarks comprise the baryons and, as a result, theoretically there should be more states compared to experimentally observed ones. One suggestion to solve this problem is considering a heavy quark-light diquark picture, which reduces the number of excited states as a result of the reduction of the number of degrees of freedom. Considering this, we adopt an interpolating current in our calculation in the form of a heavy quark-light diquark with quantum numbers $J^P=3/2^+$. In the present study, to provide further support to the results that we obtain, we also investigate the widths for $\Lambda_Q\rightarrow \Sigma^{(*)}_Q\pi$ decays of the states under consideration. In this part of the calculations, the results obtained from the mass and residue calculations are used as input parameters, and the consistency of our findings with the experimental results are checked.

This paper has the following organization. In Sec. II we give the details of the QCD sum rules calculations for the spectroscopic parameters of the considered states. In this section we also present the numerical analyses and displaying of the results for the mass sum rules. In section III, using the obtained results of the previous section, we calculate the widths for  $\Lambda_b\rightarrow \Sigma^{(*)}_b \pi$ and $\Lambda_c\rightarrow \Sigma^{(*)}_c \pi$ channels and numerically analyze the obtained sum rules. The last section contains a summary of the results and conclusions.

\section{Spectroscopic parameters of $\Lambda_b$ and $\Lambda_c$ states}

After choosing a proper  interpolating current that carries the same quantum numbers and same quark field operators in accordance with valance quark content, the following correlation function is chosen to calculate the spectroscopic parameters of the states under consideration:
\begin{equation}
\Pi _{\mu \nu}(k)=i\int d^{4}xe^{ik\cdot x}\langle 0|\mathcal{T}\{J_{\mu}(x)\bar{J}_{\nu}(0)\}|0\rangle ,  \label{eq:CorrF1}
\end{equation}        
where $\mathcal{T}$ is time ordering operator and  $J_{\mu}$ is the interpolating current with following explicit form~\cite{Wang:2017vtv}:
\begin{eqnarray}
J^{\mu}=\epsilon^{abc}[\partial_{\alpha}\partial_{\beta}u^T_aC\gamma_5 d_b+\partial_{\alpha}u^T_aC\gamma_5\partial_{\beta}d_b+\partial_{\beta}u^T_aC\gamma_5\partial_{\alpha}d_b+u^T_aC\gamma_5\partial_{\alpha}\partial_{\beta}d_b](g^{\alpha\mu}g^{\beta\delta}+g^{\alpha\delta}g^{\beta\mu}-\frac{1}{2}g^{\alpha\beta}g^{\mu\delta})\gamma^{\delta}\gamma_5Q_c.\label{Intcur}
\end{eqnarray}
In the above interpolating current, the $Q$ represents $b(c)$ quark field, $C$ is charge conjugation operator and the indices $a$, $b$ and $c$ display the colors.

One follows two paths to calculate the correlation function. In the first one, it is computed in terms of hadronic degrees of freedom. This is done by saturation of the correlation function by a complete set of hadronic states with the same quantum numbers of the interpolating current. After that the results emerge in terms of hadronic degrees of freedom such as the current coupling constant and mass of the considered hadron. This procedure leads to
\begin{eqnarray}
\Pi _{\mu \nu}^{\mathrm{Had}}(k)=\frac{\langle0|J_{\mu}|\Lambda_Q(k,s)\rangle\langle\Lambda_Q(k,s)|\bar{J}_{\nu}|0\rangle}{m_{\Lambda_Q}^2-k^2}+\cdots.
\label{cor:Phen}
\end{eqnarray}
The $\cdots$ represents the contributions of the higher states and continuum. The matrix element $\langle0|J_{\mu}|\Lambda_Q(k,s)\rangle$ in the last result is parameterized in terms of the current coupling constant, $\lambda_{\Lambda_{Q}}$, and spin vector in Rarit-Schwinger representation, $u_{\mu}(k,s)$, as
\begin{eqnarray}
\langle0|J_{\mu}|\Lambda_Q(k,s)\rangle=\lambda_{\Lambda_Q}u_{\mu}(k,s).
\end{eqnarray}  
When this matrix element is used in Eq.~(\ref{cor:Phen}) we need to perform the following summation over spin $s$:
\begin{eqnarray}
\sum_{s}u_{\mu}(k,s)\bar{u}_{\nu}(k,s)=(\not\!k+m)(-g_{\mu\nu}+\frac{\gamma_{\mu}\gamma_{\nu}}{3}+\frac{2k_{\mu}k_{\nu}}{3m_{\Lambda_Q}^2}-\frac{k_{\mu}\gamma_{\nu}-k_{\nu}\gamma_{\mu}}{3m_{\Lambda_Q}}),\label{spinsum3/2}
\end{eqnarray}
which recasts the result into the form
\begin{eqnarray}
\Pi _{\mu \nu}^{\mathrm{Had}}(k)=\frac{\lambda_{\Lambda_Q}^2(\not\!k+m_{\Lambda_Q})}{m_{\Lambda_Q}^2-k^2}(-g_{\mu\nu}+\frac{\gamma_{\mu}\gamma_{\nu}}{3}+\frac{2k_{\mu}k_{\nu}}{3m_{\Lambda_Q}^2}-\frac{k_{\mu}\gamma_{\nu}-k_{\nu}\gamma_{\mu}}{3m_{\Lambda_Q}})+\cdots.
\end{eqnarray}

The interpolating current used in the calculations couples not only with spin-$\frac{3}{2}$ states but also spin-$\frac{1}{2}$ states. Therefore to refrain from the contributions of spin-$\frac{1}{2}$ states and isolate the terms related only to spin-$\frac{3}{2}$ states, we choose a proper Lorentz structure free from spin-$\frac{1}{2}$ contribution. To this end, we consider the following matrix element showing the coupling of the chosen current to spin-$\frac{1}{2}$ states:
\begin{eqnarray}
\langle 0|J_{\mu}|\frac{1}{2}^+(k)\rangle =C_{\frac{1}{2}^+}(\gamma_{\mu}-\frac{4k_{\mu}}{m_{\frac{1}{2}^+}}) u(k,s).\label{Eq:matrixelspin1/2}
\end{eqnarray}
This matrix element indicates that the terms containing $\gamma_{\mu}$ and $k_{\mu}$ in the Lorentz structures take also contributions from spin-$\frac{1}{2}$ states due to the coupling of the current with them. To isolate the spin-$\frac{3}{2}$ states we make our analyses with the Lorentz structure $\not\!kg_{\mu\nu}$. Finally, the hadronic side results in 
\begin{eqnarray}
\tilde{\Pi}^{\mathrm{Had}}_{\mu \nu}(k)=\lambda_{\Lambda_Q}^2e^{-\frac{m_{\Lambda_Q}^2}{M^2}}\not\!kg_{\mu\nu}+\cdots,
\end{eqnarray}
after the Borel transformation. $\tilde{\Pi}^{\mathrm{Had}}_{\mu \nu}(k)$ represents the Borel transformed correlation function obtained for hadronic side, the $\cdots$ in the last result stands for both the contributions coming from the other Lorentz structures and from higher states and continuum.

The second step in the calculations is computation of the correlation function in terms of QCD degrees of freedom such as QCD condensates, quark masses and QCD coupling. To accomplish this part of the calculations, the interpolating current is used explicitly in the correlator and possible contractions between the quark fields are carried out using Wick's theorem. This turns the result into a form containing heavy and light quark propagators:
\begin{eqnarray}
\Pi^{\mathrm{QCD}}_{\mu\nu}&=&-i\int d^4xe^{ikx}\epsilon_{abc}\epsilon_{a'b'c'}\Bigg\{
\mathrm{Tr}\Big[\big[\partial^{\alpha}_{x}\partial^{\beta}_{x}\partial^{\alpha'}_{y}\partial^{\beta'}_{y}S_u^{aa'}(x-y)\big]\gamma_5 C S_d^{T,bb'}(x-y)C\gamma_5\Big]
\nonumber\\
&+&\mathrm{Tr}\Big[\big[\partial^{\alpha}_{x}\partial^{\beta}_{x}\partial^{\alpha'}_{y}S_u^{aa'}(x-y)\big]\gamma_5 C \big[\partial^{\beta'}_{y}S_d^{T,bb'}(x-y)\big]C\gamma_5\Big]+\mathrm{Tr}\Big[\big[\partial^{\alpha}_{x}\partial^{\beta}_{x}\partial^{\beta'}_{y}S_u^{aa'}(x-y)\big]\gamma_5 C\big[ \partial^{\alpha'}_{y}S_d^{T,bb'}(x-y)\big]C\gamma_5\Big]
\nonumber\\
&+&\mathrm{Tr}\Big[\big[\partial^{\alpha}_{x}\partial^{\beta}_{x}S_u^{aa'}(x-y)\big]\gamma_5 C \big[ \partial^{\alpha'}_{y} \partial^{\beta'}_{y}S_d^{T,bb'}(x-y)\big]C\gamma_5\Big]+\mathrm{Tr}\Big[\big[\partial^{\alpha}_{x}\partial^{\alpha'}_{y}\partial^{\beta'}_{y}S_u^{aa'}(x-y)\big]\gamma_5 C \big[\partial^{\beta}_{x}S_d^{T,bb'}(x-y)\big]C\gamma_5\Big]
\nonumber\\
&+&\mathrm{Tr}\Big[\big[\partial^{\alpha}_{x}\partial^{\alpha'}_{y}S_u^{aa'}(x-y)\big]\gamma_5 C \big[\partial^{\beta}_{x}\partial^{\beta'}_{y}S_d^{T,bb'}(x-y)\big]C\gamma_5\Big]+\mathrm{Tr}\Big[\big[\partial^{\alpha}_{x}\partial^{\beta'}_{y}S_u^{aa'}(x-y)\big]\gamma_5 C \big[\partial^{\beta}_{x}\partial^{\alpha'}_{y}S_d^{T,bb'}(x-y)\big]C\gamma_5\Big]
\nonumber\\
&+&\mathrm{Tr}\Big[\big[\partial^{\alpha}_{x}S_u^{aa'}(x-y)\big]\gamma_5 C \big[\partial^{\beta}_{x}\partial^{\alpha'}_{y}\partial^{\beta'}_{y}S_d^{T,bb'}(x-y)\big]C\gamma_5\Big]+\mathrm{Tr}\Big[\big[\partial^{\beta}_{x}\partial^{\alpha'}_{y}\partial^{\beta'}_{y}S_u^{aa'}(x-y)\big]\gamma_5 C \big[\partial^{\alpha}_{x}S_d^{T,bb'}(x-y)\big]C\gamma_5\Big]
\nonumber\\
&+&
\mathrm{Tr}\Big[\big[\partial^{\beta}_{x}\partial^{\alpha'}_{y}S_u^{aa'}(x-y)\big]\gamma_5 C \big[\partial^{\alpha}_{x}\partial^{\beta'}_{y}S_d^{T,bb'}(x-y)\big]C\gamma_5\Big]+\mathrm{Tr}\Big[\big[\partial^{\beta}_{x}\partial^{\beta'}_{y}S_u^{aa'}(x-y)\big]\gamma_5 C \big[\partial^{\alpha}_{x}\partial^{\alpha'}_{y}S_d^{T,bb'}(x-y)\big]C\gamma_5\Big]
\nonumber\\
&+&\mathrm{Tr}\Big[\big[\partial^{\beta}_{x}S_u^{aa'}(x-y)\big]\gamma_5 C \big[\partial^{\alpha}_{x}\partial^{\alpha'}_{y}\partial^{\beta'}_{y}S_d^{T,bb'}(x-y)\big]C\gamma_5\Big]+\mathrm{Tr}\Big[\big[\partial^{\alpha'}_{y}\partial^{\beta'}_{y}S_u^{aa'}(x-y)\big]\gamma_5 C \big[\partial^{\alpha}_{x}\partial^{\beta}_{x}S_d^{T,bb'}(x-y)\big]C\gamma_5\Big]
\nonumber\\
&+&\mathrm{Tr}\Big[\big[\partial^{\alpha'}_{y}S_u^{aa'}(x-y)\big]\gamma_5 C \big[\partial^{\alpha}_{x}\partial^{\beta}_{x}\partial^{\beta'}_{y}S_d^{T,bb'}(x-y)\big]C\gamma_5\Big]+\mathrm{Tr}\Big[\big[\partial^{\beta'}_{y}S_u^{aa'}(x-y)\big]\gamma_5 C \big[\partial^{\alpha}_{x}\partial^{\beta}_{x}\partial^{\alpha'}_{y}S_d^{T,bb'}(x-y)\big]C\gamma_5\Big]
\nonumber\\
&+&\mathrm{Tr}\Big[S_u^{aa'}(x-y)\gamma_5 C \big[\partial^{\alpha}_{x}\partial^{\beta}_{x}\partial^{\alpha'}_{y}\partial^{\beta'}_{y}S_d^{T,bb'}(x-y)\big]C\gamma_5\Big]
\Bigg\}T_{\alpha\beta\mu}S_Q^{cc'}(x-y)T_{\alpha'\beta'\nu},
\label{eqn:QCDcontracted}
\end{eqnarray}
where $\partial^{\alpha}_{x}=\frac{\partial}{{\partial x_\alpha}}$, $\partial^{\alpha'}_{y}=\frac{\partial}{\partial y_{\alpha'}}$; and  $S_q(x-y)$ and $S_Q(x-y)$ are the light and heavy quark propagators,respectively. We have also used the short-hand notation,
\begin{eqnarray}
T_{\alpha\beta\mu}=(g_{\alpha\mu}g_{\beta\delta}+g_{\alpha\delta}g_{\beta\mu}-\frac{1}{2}g_{\alpha\beta}g_{\mu\delta})\gamma^{\delta}\gamma_5.
\end{eqnarray}
In the last equation, after taking the derivatives we set $y$ to zero. The propagators in Eq.~(\ref{eqn:QCDcontracted}) are used explicitly in the calculations to obtain the QCD side of the sum rules. Their explicit forms are
\begin{eqnarray}
S_{q,}{}_{ab}(x)&=&i\delta _{ab}\frac{\slashed x}{2\pi ^{2}x^{4}}-\delta _{ab}%
\frac{m_{q}}{4\pi ^{2}x^{2}}-\delta _{ab}\frac{\langle \overline{q}q\rangle
}{12} +i\delta _{ab}\frac{\slashed xm_{q}\langle \overline{q}q\rangle }{48}%
-\delta _{ab}\frac{x^{2}}{192}\langle \overline{q}g_{\mathrm{s}}\sigma
Gq\rangle +i\delta _{ab}\frac{x^{2}\slashed xm_{q}}{1152}\langle \overline{q}%
g_{\mathrm{s}}\sigma Gq\rangle  \notag \\
&&-i\frac{g_{\mathrm{s}}G_{ab}^{\alpha \beta }}{32\pi ^{2}x^{2}}\left[ %
\slashed x{\sigma _{\alpha \beta }+\sigma _{\alpha \beta }}\slashed x\right]
-i\delta _{ab}\frac{x^{2}\slashed xg_{\mathrm{s}}^{2}\langle \overline{q}%
q\rangle ^{2}}{7776} ,  \label{eq:qprop}
\end{eqnarray}%
and 
\begin{eqnarray}
S_{Q,}{}_{ab}(x)&=&i\int \frac{d^{4}t}{(2\pi )^{4}}e^{-itx}\Bigg \{\frac{\delta
_{ab}\left( {\not\!t}+m_{Q}\right) }{t^{2}-m_{Q}^{2}}  -\frac{g_{s}G_{ab}^{\alpha \beta }}{4}\frac{\sigma _{\alpha \beta }\left( {%
\not\!t}+m_{Q}\right) +\left( {\not\!t}+m_{Q}\right) \sigma _{\alpha
\beta }}{(t^{2}-m_{Q}^{2})^{2}} +\frac{g_{\mathrm{s}}^{2}G^{2}}{12}\delta _{ab}m_{Q}\frac{t^{2}+m_{Q}{%
\not\!t}}{(t^{2}-m_{Q}^{2})^{4}} \notag \\
&&+\frac{g_{\mathrm{s}}^{3}G^{3}}{48}\delta
_{ab}\frac{\left( {\not\!t}+m_{Q}\right) }{(t^{2}-m_{Q}^{2})^{6}}  \left[ {\not\!t}\left( t^{2}-3m_{Q}^{2}\right) +2m_{Q}\left(
2t^{2}-m_{Q}^{2}\right) \right] \left( {\not\!t}+m_{Q}\right) +\ldots %
\Bigg \},  \notag \\
&&{}  \label{eq:Qprop}
\end{eqnarray}%
for the light and the heavy quark propagators in the coordinate space, respectively. The following notations are also used in Eqs.\ (\ref{eq:qprop}) and (\ref{eq:Qprop}):
\begin{eqnarray}
&&G_{ab}^{\alpha \beta }=G_{A}^{\alpha \beta
}t_{ab}^{A},\,\,~~G^{2}=G_{\alpha \beta }^{A}G_{\alpha \beta }^{A},\,\,~~G^{3}=\,\,f^{ABC}G_{\mu \nu }^{A}G_{\nu \delta }^{B}G_{\delta \mu }^{C},
\label{eq:GG}
\end{eqnarray}%
with $A,B,C=1,\,2\,\ldots 8$ and  $t^{A}=\lambda ^{A}/2$. $\lambda ^{A}$ are the Gell-Mann matrices, and the $G^A_{\alpha\beta}$ represent the gluon field strength tensors. Insertion of the propagators into the correlation function is followed by Fourier and Borel transformations. Finally, continuum subtraction is applied and the following result is achieved:
\begin{eqnarray}
\tilde{\Pi}^{\mathrm{QCD}}=\int_{(m_Q+m_u+m_d)^2}^{s_0}dse^{-\frac{s}{M^2}}\rho(s)+\Gamma.
\label{Cor:QCD}
\end{eqnarray}
where $s_0$ is the continuum threshold and  $\rho(s)$  is the spectral density obtained from the imaginary part of the  of correlation function, viz $\frac{1}{\pi}\mathrm{Im}[\Pi^{\mathrm{QCD}}]$. In the analyses, as it was stated, to isolate the contribution coming only from the spin-$\frac{3}{2}$ states the Lorentz structure is chosen as $\not\!kg_{\mu\nu}$. The standard calculations lead to the following results for $\rho(s)$ and $\Gamma$ corresponding to this Lorentz structure:
\begin{eqnarray}
\rho(s)&=&\rho_{\mathrm{Pert}}(s)+\rho_{\mathrm{Dim3}}(s)+\rho_{\mathrm{Dim4}}(s)+\rho_{\mathrm{Dim6}}(s),\nonumber\\
\end{eqnarray}
where
\begin{eqnarray}
\rho_{\mathrm{Pert}}(s)&=&-\int_0^1 dx\frac{1}{256 \pi^4 (x-1)^2} 
(m_Q^2 + s (x-1))^3 x^4 \Big[m_Q^2 (8 x-3 ) + s (3 - 19 x + 16 x^2)\Big]\theta[L(s, x)],
\nonumber\\ 
\rho_{\mathrm{Dim3}}(s)&=&-\int_0^1 dx\frac{1}{16 \pi^2} \big[m_u \big(\langle\bar{u}u\rangle - 2 \langle\bar{d}d\rangle\big) + 
   m_d \big(\langle\bar{d}d\rangle- 2 \langle\bar{u}u\rangle \big)\big] x^2 \Big[m_Q^4 (8 x-1 ) + s^2 ( x-1 )^2 ( 12 x-1 ) \nonumber\\
   &+& 2 m_Q^2 s (1 - 11 x + 10 x^2)\Big]\theta[L(s, x)]
,\nonumber\\ 
\rho_{\mathrm{Dim4}}(s)&=&-\int_0^1 dx\frac{1}{384 \pi^2 (x-1)^2} \langle\frac{\alpha_s}{\pi}G^2\rangle x^2 \Big[3 s^2 (x-1 )^4 (12 x-1 ) + m_Q^4 (-3 + 30 x - 57 x^2 + 40 x^3)\nonumber\\& +& 
 2 m_Q^2 s (3 - 39 x + 102 x^2 - 106 x^3 + 40 x^4)\Big]\theta[L(s, x)],\nonumber\\
 \rho_{\mathrm{Dim5}}(s)&=&0,
\nonumber\\
\rho_{\mathrm{Dim6}}(s)&=&\int_0^1 dx\frac{1}{12288 \pi^4 (x-1)^2}\langle g_s^3G^3\rangle x^5\Big[ 7 s (1 - 10 x + 21 x^2 - 12 x^3) -4 m_Q^2 (-6 + 5 x + 12 x^2) \Big]\theta[L(s, x)],\nonumber\\
\end{eqnarray}
and
\begin{eqnarray}
\Gamma &=&\int_0^1 dxe^{-\frac{m_Q^2}{M^2(1-x)}} \frac{1}{20480 \pi^4 (x-1)^4}\langle g_s^3G^3\rangle m_Q^4 (x-8) x^5.
\end{eqnarray}
Here $\theta[...]$ is the usual unit-step function and
\begin{eqnarray}
L(s, x)&=& s x (1 - x)-m_Q^2 x.
\end{eqnarray}

After completing the calculations for both the hadronic and QCD sides, the next stage is equating the coefficient of the same Lorentz structure obtained from each side, that is $\not\!kg_{\mu\nu}$, as a result we get
\begin{eqnarray}
\lambda_{\Lambda_Q}^2e^{-\frac{m_{\Lambda_Q}^2}{M^2}}=\tilde{\Pi}^{\mathrm{QCD}}.
\label{eq:dispersion}
\end{eqnarray}
Using this relation we obtain the masses of the considered hadrons and their current coupling constants. Thus,  for the mass we obtain
\begin{eqnarray}
m_{\Lambda_Q}^2=\frac{\frac{d}{d(-\frac{1}{M^2})}\Big[\int_{(m_Q+m_u+m_d)^2}^{s_0}dse^{-\frac{s}{M^2}}\rho(s)+\Gamma\Big]}{\Big[\int_{(m_Q+m_u+m_d)^2}^{s_0}dse^{-\frac{s}{M^2}}\rho(s)+\Gamma\Big]},
\end{eqnarray}
and the current coupling constant is obtained  as
\begin{eqnarray}
\lambda_{\Lambda_Q}^2=e^{\frac{m_{\Lambda_Q}^2}{M^2}}\Big[\int_{(m_Q+m_u+m_d)^2}^{s_0}dse^{-\frac{s}{M^2}}\rho(s)+\Gamma\Big].
\end{eqnarray}

Now, we numerically analyze the sum rules obtained  using the input parameters given in Table~\ref{tab:Param} and the working windows of auxiliary parameters such as threshold parameter $s_0$ and Borel parameter $M^2$. Although our main focus in the present work is the mass and current coupling constant of $\Lambda_b(6146)^0$ state, for completeness we also calculate the mass and current coupling constant for $\Lambda_c(2860)^+$ state. 
\begin{table}[tbp]
\begin{tabular}{|c|c|}
\hline\hline
Parameters & Values \\ \hline\hline
$m_{c}$                                     & $1.27\pm 0.02~\mathrm{GeV}$ \cite{Tanabashi2018}\\
$m_{b}$                                     & $4.18^{+0.03}_{-0.02}~\mathrm{GeV}$ \cite{Tanabashi2018}\\
$m_{u}$                                     & $2.16^{+0.49}_{-0.26}~\mathrm{MeV}$ \cite{Tanabashi2018}\\
$m_{d}$                                     & $4.67^{+0.48}_{-0.17}~\mathrm{MeV}$ \cite{Tanabashi2018}\\
$\langle \bar{q}q \rangle (1\mbox{GeV})$    & $(-0.24\pm 0.01)^3$ $\mathrm{GeV}^3$ \cite{Belyaev:1982sa}  \\
$m_{0}^2 $                                  & $(0.8\pm0.1)$ $\mathrm{GeV}^2$ \cite{Belyaev:1982sa}\\
$\langle \frac{\alpha_s}{\pi} G^2 \rangle $ & $(0.012\pm0.004)$ $~\mathrm{GeV}^4 $\cite{Belyaev:1982cd}\\
$\langle g_s^3 G^3 \rangle $                & $ (0.57\pm0.29)$ $~\mathrm{GeV}^6 $\cite{Narison:2015nxh}\\
\hline\hline
\end{tabular}%
\caption{Some input parameters used in the calculations of the masses and current coupling constants.}
\label{tab:Param}
\end{table}

To determine the working intervals for the auxiliary parameters we consider the criteria of the QCD sum rule method such as the convergence of OPE and dominance of the pole contribution. Besides these requirements, the dependencies of the results on these parameters are demanded to be relatively weak. As an asymptotic expansion, the dominant contribution to the OPE side should come from perturbative contribution and the terms with higher dimensions contribute less and less. To fix the lover limit of the Borel parameter we consider the convergence ratio, $CR(M^2)$, that is the ratio of the contribution of the highest dimensional term in the OPE side to the total one and it is given as 
\begin{eqnarray}
CR(M^2)=\frac{\Pi_{Dim6}(M^2,s_0)}{\Pi(M^2,s_0)}.
\end{eqnarray}
To determine the lover limit of Borel parameter we consider this ratio to be less than~$5~\%$  for $\Lambda_Q$ state.
The pole contribution, $PC(M^2)$ is considered to be larger or at least equal to the 10~$\%$ for the $D$-wave state,
\begin{eqnarray}
PC(M^2)=\frac{\Pi(M^2,s_0)}{\Pi(M^2,\infty)}\geq 0.10.
\end{eqnarray}
Our analyses result in the following intervals of the Borel parameters:
\begin{eqnarray}
5.2~\mathrm{GeV}^2\leq M^2 \leq 6.2~\mathrm{GeV}^2,\label{Eq:Msq1}
\end{eqnarray}
for $\Lambda_{b}$ state and 
\begin{eqnarray}
2.8~\mathrm{GeV}^2\leq M^2 \leq 3.2~\mathrm{GeV}^2,\label{Eq:Msq2}
\end{eqnarray}
for $\Lambda_{c}$ state. 
In the analyses, the working windows of the threshold parameters, $s_0$ are decided as
\begin{eqnarray}
41.5~\mathrm{GeV}^2\leq s_0 \leq 43.3~\mathrm{GeV}^2,\label{Eq:s01}
\end{eqnarray}
for $\Lambda_{b}$ state and 
\begin{eqnarray}
10.8~\mathrm{GeV}^2\leq s_0 \leq 11.6~\mathrm{GeV}^2,\label{Eq:s02}
\end{eqnarray}
for $\Lambda_{c}$ state. In these intervals the variations of the physical quantities with respect to the changes of $s_0$ are weak. The weak dependencies of the results on the auxiliary parameters form the main parts of the errors present in predictions of the QCD sum rules method. With these errors and the errors coming from the other input parameters used in the analyses our results are presented in Table~\ref{tab:results}.
\begin{table}[tbp]
\begin{tabular}{|c|c|c|}
\hline\hline
The state               & Mass (MeV) & Current coupling constant $\lambda~(\mbox{GeV}^5)$  \\
 \hline\hline
$ {\Lambda_b}$     & $6144\pm 68 $ & $ 0.264\pm0.039$ \\
$ {\Lambda_c}$     & $2855\pm 66 $ & $ 0.080\pm 0.012$ \\
\hline\hline
\end{tabular}%
\caption{The results of the masses and current coupling constants obtained for $1D$ wave $\Lambda_b$ and $\Lambda_c$ states with $J^P=\frac{3}{2}^+$.}
\label{tab:results}
\end{table}
Note that, as the interpolating currents for the $D$-wave baryons contain second order derivatives their residues or current coupling constants are obtained in $\mathrm{GeV}^5$ against the usual $S$-wave and $P$-wave baryonic states that these quantities are in $\mathrm{GeV}^3$. 

\section{The strong decays $\Lambda_Q\rightarrow \Sigma^{(*)}_Q\pi$}

 The dominant decays of $ \Lambda_b(6146)^0 $ is considered to be  
 $\Lambda_b(6146)^0\rightarrow\Sigma_b\pi$ and  $\Lambda_b(6146)^0\rightarrow\Sigma^*_b\pi$ \cite{Chen:2019ywy}.
Hence, we consider these strong decays in this section. However, the width in $ \Sigma_b\pi $ is expected to be roughly four times greater than that of the $ \Sigma^*_b\pi $ channel.   Therefore between these two channels the dominant one is $ \Sigma_b\pi $. In this accordance, we calculate the widths of the  strong decays of the $1D$-wave $\Lambda_Q$ states to $\Sigma_Q\pi$ by calculating the relevant coupling constants, $g_{\Lambda_Q\Sigma_Q\pi}$, in the framework of QCD sum rules. The calculations of the strong coupling constants are done through the following three-point correlation function:
\begin{equation}
\Pi _{\mu}(p,p')=i^2\int d^{4}xe^{-ip\cdot x}\int d^{4}ye^{ip'\cdot y}\langle 0|\mathcal{T}\{J_{\Sigma_Q}(y)J_{\pi}(0)\bar{J}_{\mu}(x)\}|0\rangle ,  \label{eq:CorrF2}
\end{equation}    
 where $J_{\mu}(x)$ is the interpolating current given in Eq.~(\ref{Intcur}) for the $\Lambda_Q$ state under consideration. The interpolating currents for the spin-$\frac{1}{2}$ $\Sigma_Q$  and pion states are as follows:
 \begin{eqnarray}
 J_{\Sigma_Q}&=&\epsilon_{ijk}(u^{iT}C\gamma^{\alpha}u^{j})\gamma_5 \gamma_{\alpha}Q^k,\nonumber\\ 
J_{\pi}&=&i\bar{u}^{l}\gamma_5 d^{l},\label{currents}
 \end{eqnarray}
where $i,~j,~k$ and $l$ are the color indices, $T$ represents transpose.

As in the mass calculation, for the strong coupling constant calculation, we follow a similar procedure and compute the correlator in terms of QCD degrees of freedom on one side and hadronic degrees of freedom on the other side. 

We insert complete sets of hadronic states into the correlator to deduce the result in terms of hadronic degrees of freedom. This part results in 
\begin{eqnarray}
\Pi_{\mu}^{Had}(p,p')=\frac{\langle 0|J_{\Sigma_Q}|\Sigma_Q(p',s')\rangle \langle0|J_{\pi}|\pi(q)\rangle\langle\pi(q)\Sigma(p',s')|\Lambda_Q(p,s)\rangle\langle\Lambda_Q(p,s)|J_{\mu}|0\rangle}{(m_{\Lambda_Q}^2-p^2)(m_{\Sigma_Q}^2-p'^2)(m_{\pi}^2-q^2)}+\cdots.\label{corrdecay}
\end{eqnarray}   
The $\cdots$ in Eq.~(\ref{corrdecay}) is used to represent the contribution of the higher states and continuum. The matrix elements present in this result are parametrized in terms of physical parameters as follows:
\begin{eqnarray}
\langle 0|J_{\Sigma_Q}|\Sigma_Q(p',s')\rangle &=&\lambda_{\Sigma_Q}u(p',s'),\nonumber\\
\langle0|J_{\pi}|\pi(q)\rangle &=& i\frac{f_{\pi}m_{\pi}^2}{(m_u+m_d)},\nonumber\\
\langle0|J_{\mu}|\Lambda_Q(p,s)\rangle &=&\lambda_{\Lambda_Q}u_{\mu}(p,s),
\end{eqnarray}
and the following matrix element is defined in terms of the considered strong coupling constant $g_{\Lambda_{Q}\Sigma_Q\pi}$ as
\begin{eqnarray}
\langle\pi(q)\Sigma_Q(p',s')|\Lambda_Q(p,s)\rangle = g_{\Lambda_{Q}\Sigma_Q\pi} \bar{u}(p',s')u_{\nu}(p,s)q^{\nu}.
\end{eqnarray} 
When we use these relations in Eq.~(\ref{corrdecay}) the final form of the correlator in the physical side becomes
\begin{eqnarray}
\Pi_{\mu}^{Had}(p,p')&=-i&\frac{f_{\pi}m_{\pi}^2}{(m_u+m_d)}\frac{\lambda_{\Sigma_Q}\lambda_{\Lambda_Q}^{*} g_{\Lambda_{Q}\Sigma_Q\pi}} {(m_{\Lambda_Q}^2-p^2)(m_{\Sigma_Q}^2-p'^2)(m_{\pi}^2-q^2)}(\not\!p'+m_{\Sigma_Q})\nonumber\\
&\times & (\not\!p+m_{\Lambda_Q})(-g_{\nu\mu}+\frac{\gamma_{\nu}\gamma_{\mu}}{3}+\frac{2p_{\nu}p_{\mu}}{3m_{\Lambda_Q}^2}-\frac{p_{\nu}\gamma_{\mu}-p_{\mu}\gamma_{\nu}}{3m_{\Lambda_Q}})q^{\nu}.\label{corrdecay1}
\end{eqnarray}   
To obtain the last result we use the Eq.~(\ref{spinsum3/2}) together with the following summation over spins of Dirac spinors:
\begin{eqnarray}
\sum_{s'}u(p',s')\bar{u}(s',p')=(\not\!p'+m_{\Sigma_Q}).
\end{eqnarray}
To suppress the contribution of the higher states and continuum we apply double Borel transformation and obtain the final form of the result for this side as
\begin{eqnarray}
\mathcal{B}\Pi_{\mu}^{Had}(p,p')&=&-i\frac{f_{\pi}m_{\pi}^2\lambda_{\Sigma_Q}\lambda_{\Lambda_Q}^* g_{\Lambda_{Q}\Sigma_Q\pi}e^{-\frac{m_{\Lambda_Q}^2}{M^2}}e^{-\frac{m_{\Sigma_Q}^2}{M'^2}}} {(m_u+m_d)(m_{\pi}^2-q^2)}\Big\{\frac{(m_{\Lambda_Q}^2 - m_{\Lambda_Q} m_{\Sigma_Q} + m_{\Sigma_Q}^2 - q^2)}{3 m_{\Lambda_Q}^2}\not\!q \not\!p' q_{\mu}+\cdots\Big\}.\label{corrdecay2}
\end{eqnarray}   
In the last result, we only give the term that we use in the analyses and there are also other Lorentz structures giving contributions. As we mentioned already, the interpolating current of spin-$\frac{3}{2}$ states also couples to the spin-$\frac{1}{2}$ states. In this part, in order to focus the contribution of only the spin-$\frac{3}{2}$ states, we make a special ordering in the Dirac matrices. Considering the matrix element given in Eq.~(\ref{Eq:matrixelspin1/2}) it can be seen that, the terms taking the contribution from spin-$\frac{1}{2}$ states are related to the Lorentz structures containing $\gamma_{\mu}$ or $(p'+q)_{\mu}$ at the far right end. Therefore the Dirac matrices are ordered in the form $\not\!q \not\!p'\gamma_{\mu}$ first and  then the structure, $\not\!q \not\!p' q_{\mu}$, ruling out the spin-$\frac{1}{2}$ contributions, is chosen.    The contributions of other structures and the contributions coming from higher states are represented by $\cdots$. $M^2$ and $M'^2$ in Eq.(\ref{corrdecay2}) are the Borel parameters.

The other side of the calculation requires to use interpolating currents given in Eqs.~(\ref{Intcur}) and (\ref{currents}) explicitly inside the correlator. Possible contractions between the quark fields performed via Wick's theorem render the result into the form containing heavy and light quark propagators. Their explicit expressions, Eqs.~(\ref{eq:qprop}) and (\ref{eq:Qprop}), are used and correlation function $\Pi_{\mu }^{QCD}(p,p')$ is obtained with different Dirac structures as in the hadronic side. The results are lengthy and to refrain overwhelming long expressions we shortly represent the result here as
\begin{eqnarray}
\Pi_{\mu }^{QCD}(p,p')&=&\Pi^{QCD}_{1}(q^2)\, \not\!q \not\!p' q_{\mu}
+\mathrm{other \,\,\, structures}, \label{eq:PiOPE}
\end{eqnarray}
and not give their explicit form. The invariant function $\Pi^{QCD}_{1}(q^2)$ here is the coefficient of the  $\not\!q \not\!p' q_{\mu}$ structure that we use in the analyses. The imaginary parts of the obtained results are used as spectral densities in the following dispersion integral leading us to the final form of the QCD side 
\begin{eqnarray}
\Pi^{QCD}_{i}(q^2)=\int ds\int
ds'\frac{\rho_{i}^{pert}(s,s',q^{2})+\rho_{i}^{non-pert}(s,s',q^{2})}{(s-p^{2})
(s'-p'^{2})}. \label{eq:Pispect}
\end{eqnarray}
where $i$ represents different Lorentz structures present in the calculation and the spectral densities are represented by their perturbative and nonperturbative parts as $ \rho_{i}^{pert}(s,s',q^{2}) $ and $ \rho_{i}^{non-pert}(s,s',q^{2}) $, respectively.

The results obtained from the hadronic and the QCD sides are matched, considering the same Lorentz structure, giving us the QCD sum rules for the strong coupling constants under question as follows:
\begin{eqnarray}
g_{\Lambda_{Q}\Sigma_Q\pi}(Q^2)=-i\frac{3m_{\Lambda_Q}^2(m_{\pi}^2+Q^2)e^{\frac{m_{\Lambda_Q}^2}{M^2}}e^{\frac{m_{\Sigma_Q}^2}{M'^2}}}{f_{\pi} \lambda_{\Lambda_Q}^*\lambda_{\Sigma_Q}\mu_{\pi}(m_{\Lambda_Q}^2 - m_{\Lambda_Q} m_{\Sigma_Q} + m_{\Sigma_Q}^2 +Q^2)}\mathcal{B}\Pi_1^{QCD}(s,s',Q^2), \label{eq:coupling}
\end{eqnarray}
where $\mathcal{B}\Pi_1^{QCD}(s,s',Q^2)$ is the result of QCD side after Borel transformation,  $Q^2=-q^2$ and $\mu_{\pi}=\frac{m_{\pi}^2}{(m_u+m_d)}$.

Now, we present the numerical computations of the coupling constants $g_{\Lambda_{Q}\Sigma_Q\pi}$ obtained for the decays $\Lambda_Q\rightarrow \Sigma^{(*)}_Q\pi$. To this end, we use  the results that we obtained from the mass analyses of $\Lambda_Q$ states as inputs.  Besides, we also need the values of some other parameters which are $\lambda_{\Sigma_b}=0.062\pm0.018~\mathrm{GeV}^3$~\cite{Azizi:2008ui}, $m_{\Sigma_b}=5810.56\pm 0.25~\mathrm{MeV}$~\cite{Tanabashi2018}, $\lambda_{\Sigma_c}=0.045\pm0.015~\mathrm{GeV}^3$~\cite{Azizi:2008ui}, $m_{\Sigma_c}=2453.97\pm 0.14~\mathrm{MeV}$~\cite{Tanabashi2018}, $f_{\pi}=131.5~\mathrm{MeV}$ and $\mu_{\pi}=-\frac{2\langle\bar{q}q\rangle}{f_{\pi}^2}$.

As for the four additional auxiliary parameters, the Borel parameter $M^2$ and the threshold parameter $s_0$ are used as in the mass sum rule calculations, Eqs.~(\ref{Eq:Msq1}), (\ref{Eq:Msq2}), (\ref{Eq:s01}) and (\ref{Eq:s02}). The second Borel parameter $M'^2$ and the second threshold parameter $s'_0$ are determined, considering the standard criteria of the QCD sum rule that we explained in mass sum rule calculations, as
\begin{eqnarray}
5.0~\mathrm{GeV}^2\leq M'^2 \leq 6.0~\mathrm{GeV}^2,\label{Eq:M'sq1}
\end{eqnarray} 
\begin{eqnarray}
39.0~\mathrm{GeV}^2\leq s'_0 \leq 41.0~\mathrm{GeV}^2,\label{Eq:s'01}
\end{eqnarray}
for the bottom baryon case and 
\begin{eqnarray}
2.7~\mathrm{GeV}^2\leq M'^2 \leq 3.1~\mathrm{GeV}^2,\label{Eq:M'sq2}
\end{eqnarray} 
\begin{eqnarray}
10.6~\mathrm{GeV}^2\leq s'_0 \leq 11.4~\mathrm{GeV}^2,\label{Eq:s'02}
\end{eqnarray}
for the charmed baryon case.

Using the related input parameters and working intervals of the auxiliary parameters, we attain the results of the coupling constants as a function of $Q^2$ which is well represented by following fit function
\begin{eqnarray}
g_{\mathrm{fit}}(Q^2)&=& c_1e^{-\frac{Q^2}{c_2}}+c_3.
\end{eqnarray}
The parameters of the fit function, $c_1$, $c_2$, and $c_3$ are determined from our analyses and presented in Table~\ref{tab:fitfunction}.
\begin{table}[tbp]
\begin{tabular}{|c|c|c|c|}
\hline\hline
The decay mode     & $c_1~(\mathrm{GeV}^{-1})$ & $c_2$ $(\mbox{GeV}^2)$ & $c_3~(\mathrm{GeV}^{-1})$\\
 \hline\hline
$ {\Lambda_b}\rightarrow \Sigma_b\pi$     & $-6734.9 $ & $7.9  $ &$6804.3 $        \\
$ {\Lambda_c}\rightarrow \Sigma_c\pi$     & $-2087.9$  & $ 4.4 $ &$2156.9 $       \\
\hline\hline
\end{tabular}%
\caption{The parameters of the fit function.}
\label{tab:fitfunction}
\end{table}
Using the fit functions of related decays, we obtain the considered coupling constants at $Q^2=-m_{\pi}^2$ for both decay channels. The results of these coupling constants are presented in Table~\ref{tab:resultofdecay}.
\begin{table}[tbp]
\begin{tabular}{|c|c|c|}
\hline\hline
The decay mode     & $g_{\Lambda_Q\Sigma_Q\pi}(\mbox{GeV}^{-1})$ & $\Gamma$ $(\mbox{MeV})$ \\
 \hline\hline
$ {\Lambda_b}\rightarrow \Sigma_b\pi$     & $52.8\pm 4.7 $ & $2.3 \pm 0.4$       \\
$ {\Lambda_c}\rightarrow \Sigma_c\pi$     & $59.6\pm 5.4 $ & $34.9\pm6.5 $       \\
\hline\hline
\end{tabular}%
\caption{The coupling constants and the calculated partial widths for considered decays.}
\label{tab:resultofdecay}
\end{table}
This table also shows the results obtained for the partial widths of considered decays of the $\Lambda_Q$ states which are calculated by applying the following equation:
\begin{eqnarray}
\Gamma (\Lambda _{Q} &\rightarrow &\Sigma _{Q}\pi)=\frac{%
g_{\Lambda _{Q}\Sigma _{Q}\pi}^{2}}{24\pi m_{\Lambda_Q}^2}\left[
(m_{\Lambda_Q}-m_{\Sigma _{Q}})^{2}-m_{\pi}^{2}\right] 
 f^{3}(m_{\Lambda_Q },m_{\Sigma_Q },m_{\pi}),\label{Eq:Width}
\end{eqnarray}
where $f(m_{\Lambda_Q },m_{\Sigma_Q },m_{\pi})$ is defined through the following expression
\begin{equation}
f(x,y,z)=\frac{1}{2x}\sqrt{%
x^{4}+y^{4}+z^{4}-2x^{2}y^{2}-2x^{2}z^{2}-2y^{2}z^{2}}.
\end{equation}
The errors in these results arise from the uncertainties of the input and auxiliary parameters.  
In Ref. \cite{Chen:2019ywy}, the following relation between the partial widths of the  $\Lambda_b(6146)^0\rightarrow\Sigma^*_b\pi$ and  $\Lambda_b(6146)^0\rightarrow\Sigma_b\pi$ strong decays in p- and f-wave decays is obtained:
\begin{equation}
\frac{\Gamma[\Lambda_b(6146)^0\rightarrow\Sigma^*_b\pi]}{\Gamma[\Lambda_b(6146)^0\rightarrow\Sigma_b\pi]}=\frac{0.65^p+0.28^f}{3.25}=0.286.
\end{equation}
We assume that this is roughly holds in c channel, as well. Using this relation, we estimate the widths in $\Sigma^*_Q\pi  $ channels and also the total widths of the states under study: all of these widths are presented in Table ~\ref{tab:resultofdecay11}.
\begin{table}[tbp]
\begin{tabular}{|c|c|c|}
\hline\hline
The decay mode      & $\Gamma$ $(\mbox{MeV})$ \\
 \hline\hline
$ {\Lambda_b}\rightarrow \Sigma_b\pi$    & $2.3 \pm 0.4$       \\
$ {\Lambda_b}\rightarrow \Sigma^*_b\pi$    & $0.7 \pm 0.1$       \\
Total& $ 3.0\pm0.4 $\\
\hline
$ {\Lambda_c}\rightarrow \Sigma_c\pi$     & $34.9\pm6.5 $       \\
$ {\Lambda_c}\rightarrow \Sigma^*_c\pi$    & $10.0 \pm 1.9$       \\
Total&$ 44.9\pm6.8  $\\
\hline\hline
\end{tabular}%
\caption{Partial and total widths for the decays under study.}
\label{tab:resultofdecay11}
\end{table}

\section{Discussion and Conclusion}

We calculated the mass and the current coupling constant of the recently observed $\Lambda_b(6146)^0$ state assigning its quantum numbers as $J^P=\frac{3}{2}^+$. This state  together with the $\Lambda_b(6152)^0$  (probably a $1D$-wave state with $J^P=\frac{5}{2}^+$) form a  $\Lambda_b(1D)^0$ doublet~\cite{Capstick:1986bm,Chen:2014nyo,Aaij:2019amv}. Based on the provided information by recent experimental results, we chose a $D$-wave type interpolating current for $\Lambda_b(6146)^0$ state. For completeness, we also calculated the spectroscopic parameters of its charmed partner $\Lambda_c$ state with the same quantum numbers. The result for the mass of the $\Lambda_b$ state was obtained to be $m_{\Lambda_b} = (6144\pm 68)$~MeV, which is in a good  consistency with other theoretical predictions: $m_{\Lambda_b} =6147$~MeV~\cite{Chen:2014nyo}, $m_{\Lambda_b} =6190$~MeV~\cite{Ebert:2011kk}, $m_{\Lambda_b}= 6181$~MeV~\cite{Roberts:2007ni}, $m_{\Lambda_b}= 6145$~MeV~\cite{Capstick:1986bm}, $m_{\Lambda_b}= 6149$~MeV~\cite{Chen:2019ywy}, and  $6.01^{+0.20}_{-0.12}$~GeV~\cite{Chen:2016phw}. Our result on the mass of the $\Lambda_b$ is in accord with the experimental data of the LHCb Collaboration, as well. This leads us to consider the $\Lambda_b(6146)^0$  state as a $1D$-wave resonance with quantum numbers $J^P=\frac{3}{2}^+$. 

The mass result obtained for $1D$ wave $\Lambda_c$ state with $J^P=\frac{3}{2}^+$ is $m_{\Lambda_c}=(2855\pm 66)$~MeV, which is also consistent, within the errors, with the predictions of Refs.~\cite{Chen:2014nyo,Ebert:2011kk,Roberts:2007ni,Capstick:1986bm,Chen:2016iyi,Chen:2016phw,Wang:2017vtv}  given as $m_{\Lambda_c}= 2857$~MeV, $m_{\Lambda_c}= 2874$~MeV, $m_{\Lambda_c}= 2887$~MeV, $m_{\Lambda_c}=2910$~MeV, $m_{\Lambda_c}=2843$~MeV, $m_{\Lambda_c}=2.81^{+0.33}_{-0.18}$~GeV, and $m_{\Lambda_c}=2.83^{+0.15}_{-0.24}$~GeV, respectively. Our result is also in agreement with experimentally observed mass value for $\Lambda_c(2860)^+$ state which is $m_{\Lambda_c(2860)^+}=2856.1^{+3.6}_{-7.8}$~MeV \cite{Aaij:2017vbw}. This can be considered as another support to assign these states as resonances in $b$ and $c$ $1D$-wave channels with spin-parity $J^P=\frac{3}{2}^+$. 

 To make a final decision on the structure and quantum numbers of these states especially the lesser-known $\Lambda_b(6146)^0$ resonance, we need to support these assignments by the width calculations, which require the calculations of the partial widths of the dominant decays of this state. The $\Lambda_b(6146)^0$ state was seen in $\Lambda_b^0\pi^+\pi^-$ channel  by LHCb collaboration \cite{Aaij:2019amv} and very recently confirmed by the CMS collaboration in the same channel \cite{Sirunyan:2020gtz}. The dominant decays of this state is considered to be  
 $\Lambda_b(6146)^0\rightarrow\Sigma_b\pi$ and  $\Lambda_b(6146)^0\rightarrow\Sigma^*_b\pi$ \cite{Chen:2019ywy}. Although, by considering the $\Lambda_b(6146)^0$  sate as a $1D$-wave resonances with quantum numbers $J^P=\frac{3}{2}^+$ and its decays to $\Sigma_b\pi$ and  $\Sigma^*_b\pi$ final states, the obtained total width via the quark potential model in this study is comparable  with the experimental data within the presented errors, the LHCb collaboration could not find significant signals in these channels \cite{Aaij:2019amv}.
 
  We considered the  $\Lambda_b(6146)^0\rightarrow\Sigma_b\pi$ and  $\Lambda_b(6146)^0\rightarrow\Sigma^*_b\pi$ decay modes. The partial width of the decay in $ \Sigma_b\pi $ channel is considered to be roughly four times greater than that of the $ \Sigma^*_b\pi $ channel \cite{Chen:2019ywy}. Hence, by calculation of the related strong coupling constant, $g_{\Lambda_b\Sigma_b\pi}$ via the three-point QCD sum rule approach in details, we obtained the  partial width of this decay as $\Gamma[\Lambda_b(6146)^0\rightarrow\Sigma_b\pi]=2.3 \pm 0.4~\mathrm{MeV} $. The partial width for this mode is obtained as $\Gamma[\Lambda_b(6146)^0\rightarrow\Sigma_b\pi]=3.25~\mathrm{MeV} $ in Ref.  \cite{Chen:2019ywy}.  In literature, there are other works on the strong decays of $1D$-wave $\Lambda_Q$ states~\cite{Yao:2018jmc,Liang:2019aag}, as well.  In these works the results $\Gamma(\Lambda_b\rightarrow\Sigma_b\pi)=4.57^{+1.09}_{-1.20}~\mathrm{MeV}$~\cite{Yao:2018jmc} and $\Gamma(\Lambda_b\rightarrow\Sigma_b\pi)=1.79~\mathrm{MeV}$~\cite{Liang:2019aag} are obtained.  As is seen, the results of Refs.~\cite{Yao:2018jmc} and  \cite{Chen:2019ywy} are considerably larger and that of the \cite{Liang:2019aag} is considerably smaller than our prediction.  We considered the ratio $ \frac{\Gamma[\Lambda_b(6146)^0\rightarrow\Sigma^*_b\pi]}{\Gamma[\Lambda_b(6146)^0\rightarrow\Sigma_b\pi]} $ obtained in Ref. \cite{Chen:2019ywy}, to estimate the partial width of $\Lambda_b(6146)^0\rightarrow\Sigma^*_b\pi$ as $ \Gamma[\Lambda_b(6146)^0\rightarrow\Sigma_b\pi]=0.7 \pm 0.1~\mathrm{MeV} $, as well. We also obtained the total width of $ \Lambda_b(6146)^0 $ state as $\Gamma_{\Lambda_{b}(6146)^0}=3.0\pm0.4$~MeV, which is in a nice consistency with the experimental data of LHCb collaboration: $\Gamma_{\Lambda_{b}(6146)^0}=2.9\pm1.3\pm0.3$~MeV.
  
  For the c-partner, we obtained $\Gamma[ {\Lambda_c}\rightarrow \Sigma_c\pi]=34.9\pm6.5 $ and $ \Gamma[{\Lambda_c}\rightarrow \Sigma^*_c\pi]= 10.0 \pm 1.9$, which leads to $\Gamma_{\Lambda_c(2860)^+}=44.9\pm6.8$~MeV for the total width of $\Lambda_c(2860)^+$ state. This result, within the errors, is consistent with the experimental data, $ \Gamma_{\Lambda_c(2860)^+}=67.6^{+17.4}_{-29.5}$~MeV \cite{Aaij:2017vbw}, as well. 
  
  Considering the mass and the obtained width from its dominant decays, the newly observed $\Lambda_b(6146)^0$  state was assigned as a $1D$-wave excited state in usual three-quark $ \Lambda_b $ channel with spin-parity $J^P=\frac{3}{2}^+$. We also assigned the same quantum numbers for its c-partner.  More experimental and theoretical effort are needed to clarify the decay modes of  $\Lambda_b(6146)^0$ and $\Lambda_c(2860)^+  $  sates  in order to more clarify their nature.






\label{sec:Num}

\end{document}